# mod_oai: An Apache Module for Metadata Harvesting

Michael L. Nelson[1], Herbert Van de Sompel[2], Xiaoming Liu[2], Terry L. Harrison[1], Nathan McFarland[2]

[1]Old Dominion University, Department of Computer Science, Norfolk VA 23508 USA
{mln, tharriso}@cs.odu.edu

[2]Los Alamos National Laboratory, Research Library, Los Alamos NM 87545 USA
{herbertv, liu_x, nmcfarl}@lanl.gov

**Abstract.** We describe mod_oai, an Apache 2.0 module that implements the Open Archives Initiative Protocol for Metadata Harvesting (OAI-PMH). OAI-PMH is the de facto standard for metadata exchange in digital libraries and allows repositories to expose their contents in a structured, application-neutral format with semantics optimized for accurate incremental harvesting. Current implementations of OAI-PMH are either separate applications that access an existing repository, or are built-in to repository software packages. mod_oai is different in that it optimizes harvesting web content by building OAI-PMH capability into the Apache server. We discuss the implications of adding harvesting capability to an Apache server and describe our initial experimental results accessing a departmental web site using both web crawling and OAI-PMH harvesting techniques.

## 1 INTRODUCTION

Considerable attention has been given to increasing the efficiency and scope of web crawlers. Commercial web crawlers are estimated to index only approximately 16% of the total "surface web" [14] and the size of the "deep web" or "hidden web" is estimated to be up to 550 times as large as the surface web [4]. These problems are due in part to the extremely large scale of the web. To increase efficiency, a number of techniques have been proposed such as more accurately estimating web page creation and updates [16, 7] and more efficient crawling strategies [6]. Techniques such as probing search engines with keyword queries are used to increase the scope of web crawls and obtain more of the deep web [12, 17, 18]. Extending the scope of a web crawl has implications on the coverage of search engines and in web preservation [10,15].

All of these approaches stem from the fact that http does not provide semantics to allow web servers to answer questions of the form "what resources do you have?" and "what resources have changed since 2004-12-27?" A number of approaches have been suggested to add update semantics to http servers, including conventions about how to store indexes as well-known URLs for crawlers [5] and a combination of indexes and http extensions [23]. WebDAV [8] provides some update semantics through http extensions, but it is not widely implemented. The RSS [19] family of syndication formats are widely implemented, but these formats are either in the process of standardization or optimized for syndicating web ephemera and do not provide selective or incremental of resources. The Open Archives Initiative Protocol for Metadata Harvesting (OAI-PMH) [13] has a very powerful and general set of update semantics and is the de facto standard for metadata interchange within the digital library community. Packages for implementing OAI-PMH repositories for XML files have been described [11, 20], but they are focused on highly constrained scenarios, not general web content and they do not integrate directly into the web server. mod_oai is an Apache module that implements OAI-PMH functionality directly into the Apache web server. The goal of this project is not to replace specific OAI-PMH repository implementations, but rather to bring OAI-PMH's more efficient update semantics to the general web crawling community.

## 2 OAI-PMH

OAI-PMH 2.0 is a low-barrier, HTTP-based protocol designed to allow incremental harvesting of XML metadata. An OAI-PMH repository is a network accessible server that can process the six OAI-PMH protocol requests, and respond to them as specified by the protocol document. A harvester is an application that issues OAI-PMH protocol requests, in order to harvest XML metadata. The OAI-PMH is based on a data model that helps in specifying the semantics of the six protocol requests. In what follows, OAI-PMH entities are written in *italic*, while OAI-PMH protocol requests are written in `courier`:

- An OAI-PMH repository exposes *metadata* about resources. By definition, resources themselves are outside of the scope of the OAI-PMH.
- The *item* is the entry point to all available *metadata* pertaining to a resource. In the protocol, the *item* is uniquely identified by an OAI-PMH *identifier*.
- An *item* can gives access to one or more *records*. *Records* contain *metadata* (and secondary information about that *metadata*). A specific *record* in the OAI-PMH is unambiguously identified by means of the combination of the OAI-PMH *identifier* (of the *item*), the *metadataPrefix* that specifies the *metadata format* used for the dissemination of the *metadata*, and the OAI-PMH *datestamp* of the *metadata*. The *datestamp* is the date and time of creation or modification of *metadata*. Note that the *datestamp* is a property of the *metadata* record, not of the *item* as used to be in OAI-PMH version 1.X. This reflects the fact that *metadata* of various *metadata formats* may be made available and may be modified independently, thus having different *datestamps*.
- The OAI-PMH also defines a *set* as an optional construct for grouping *items* for the purpose of selective harvesting. Repositories may organize *items* into *sets*. *set* organization may be flat, i.e. a simple list, or hierarchical. Multiple, parallel, *set* structures may exist.

The OAI-PMH defines three supporting protocol requests that are aimed at helping a harvester understand the nature of an OAI-PMH Repository:

- `Identify`: this verb is used to retrieve information about a repository such as administrator, harvesting granularity, etc.
- `ListMetadataFormats`: this verb is used to retrieve the *metadata formats* available from a repository.
- `ListSets`: This verb is used to retrieve the *set* structure of a repository. This information is useful for selective harvesting.

The OAI-PMH defines 3 further protocol requests that are aimed at the actual harvesting of XML metadata:

- `ListRecords`: this verb is used to harvest *records* from a repository. Optional arguments permit selective harvesting of *records* based on *set* membership and/or *datestamp*.
- `GetRecord`: This verb is used to retrieve an individual *record* from a repository. Required arguments specify the *identifier* of the *item* from which the *record* is requested and the *metadata format* of the metadata that should be included in the *record*.
- `ListIdentifiers`: This verb is an abbreviated form of `ListRecords`, retrieving only *identifiers*, *datestamps* and *set* information.

For example, for an OAI-PMH repository at baseURL http://www.arxiv.org/oai2/, the following protocol request is issued to obtain metadata in Dublin Core format for all items that have changed since December 27th 2004:

http://www.arxiv.org/oai2?verb=ListRecords&metadataPrefix=oai_dc&from=2004-12-27

Due to its origins in the realm of resource discovery, the OAI-PMH mandates the support of the Dublin Core [24] metadata format, but strongly encourages supporting more expressive formats. As a result, any metadata format can be used as long as it is defined by means of an XML Schema.

OAI-PMH uses an opaque data structure called a *resumptionToken* to separate long responses into many shorter responses. For example, if a `ListRecords` response is 1M records, neither the repository nor the harvester could likely handle that response. The repository might choose to separate the complete list into 1000 incomplete lists of 1000 records each. The distinguishing characteristic is that the repository chooses the size of the *resumptionToken*, not the harvester. This allows repositories to throttle the load placed on them by harvesters

## 3 mod_oai

mod_oai is an Apache module that responds to OAI-PMH requests on behalf of a web server. If Apache and mod_oai are installed at http://www.foo.edu/, then the baseURL for OAI-PMH requests is http://www.foo.edu/mod_oai. mod_oai exposes the files on an Apache web server as an OAI-PMH repository with the following characteristics:

- OAI-PMH data model:
  - OAI-PMH *identifier*: The URL of the resource serves as the OAI-PMH *identifier*. This choice facilitates a harvesting strategy whereby `ListIdentifiers` (with `from` and `until` parameters) is used to determine the URLs of web resources that have been updated since a previous harvest.
  - OAI-PMH *datestamp*: The modification time of the resource is used as the OAI-PMH datestamp of all 3 metadata formats. This is because all 3 metadata formats are dynamically derived from the resource itself. As a result, an update to a resource will result in new *datestamps* for all *metadata formats*.
  - OAI-PMH *sets*: A *set* organization is introduced based on the MIME type of resource. This choice facilitates MIME type specific resource harvesting, through the use of the `set` argument in protocol requests.
- Three parallel *metadata formats* are supported:
  - oai_dc: Dublin Core is supported as mandated, but only technical metadata that can be derived from http header information (file size, MIME type, etc.) is included.
  - http_header: A new *metadata format*, http_header, is introduced. It contains all http response headers that would have been returned if a web resource were obtained by means of an http GET.
  - oai_didl: This *metadata format* is introduced to allow harvesting of the resource itself. It is discussed in more detail in Section 4. In this *metadata format*, the web resource is represented by means of an XML wrapper document that is compliant with the MPEG-21 Digital Item Declaration Language (DIDL) [1], which has been devised to facilitate the representation of complex digital objects. This XML wrapper document includes the http_header, as well as the web resource itself, provided using the By-Reference or By-Value approach, or both. Figure 1 shows a structural view of a web resource represented in the oai_didl metadata format.

The aforementioned design choices allow two main classes of mod_oai use, both of which offers selective harvesting semantics with both *datestamp* and *sets* (i.e. MIME types) as selection criteria:

- Use of `ListIdentifiers` to identify URLs of web resources available from an Apache server, and using the resulting list of URLs as seeds for a regular web crawl.
- Use of `ListRecords` to harvest the web resources represented by means of XML wrapper documents compliant with MPEG-21 DIDL. To ensure that harvesters that choose this approach instead of regular crawling obtain all the information they require, http header information represented using the http_header metadata format is also included in this XML wrapper.

## 4 mod_oai and MPEG-21 DIDL

Typically, OAI-PMH repositories expose descriptive metadata about resources such as Dublin Core or MARCXML. However, recently, interest has increased in using the OAI-PMH to harvest the resources themselves, not just metadata about the resources. A technique has been proposed [21] to enable resource harvesting that is based on representing resources using an XML-based complex object format such as the MPEG-21 Digital Item Declaration Language (MPEG-21 DIDL) or the Metadata Encoding and Transmission Standard (METS). In mod_oai, this technique is used to enable OAI-PMH harvesting of Web resources, and MPEG-21 DIDL is used as the complex object format to represent the resources.

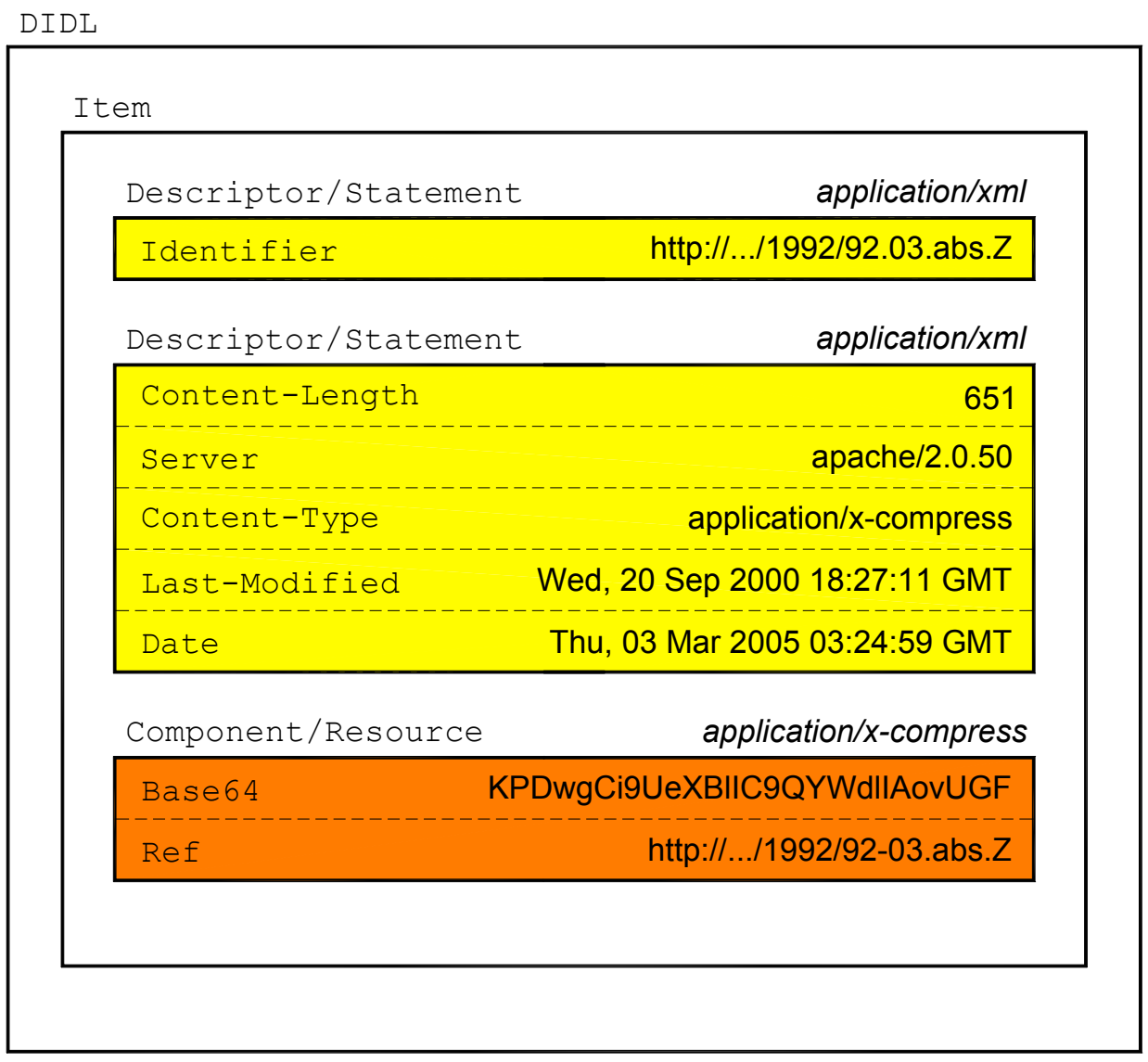

**Fig 1.** A structural view of a web resource in MPEG-21 DIDL.

MPEG-21 DIDL is an XML-based instantiation of the data model defined by the MPEG-21 Digital Item Declaration (MPEG-21 DID) ISO Standard [22], which itself is representation independent. MPEG-21 DID introduces a set of abstract concepts that, together, form a well-defined Abstract Model for declaring Digital Objects. A simplified explanation of the MPEG-21 DID Abstract Model is given here; interested readers are referred to [3] for more detailed information. The MPEG-21 DID Abstract Model recognizes several MPEG-21 DID entities (written in *italic* font style), each of which has a corresponding XML element in the DIDL XML Schema [http://purl.lanl.gov/STB-RL/schemas/2004-11/DIDL.xsd]:

- An *item* is the declarative representation of a Digital Object. It is a grouping of *items* and/or *components*.
- A *component* is a grouping of *resources*. Multiple *resources* in the same *component* are considered bit-equivalent and consequently it is left to an agent to select which one to use.

- A *resource* is an individual datastream.
- A *container* is a grouping of *containers* and/or *items*.
- Secondary information pertaining to a *container*, an *item*, or a *component* can be conveyed by means of a *descriptor/statement* construct.

When mod_oai exposes web resources via the OAI-PMH, it maps that resource occurs to an XML-based representation of the resource that is compliant with MPEG-21 DIDL. The mapping is illustrated by Figure 1 and is described here. MPEG-21 DIDL XML elements have the same name as their corresponding entity of the MPEG-21 DID data model, and, for clarity, are shown in the `courier` font:

- The Web resource is considered a Digital Object, and hence is mapped to a top-level DIDL `Item` element. Two `Descriptor/Statement` constructs are attached to this `Item` to convey secondary information pertaining to the Web resource:
  - The URI of the Web resource is provided in a `Descriptor/Statement` construct, the content of which is compliant with the MPEG-21 Digital Item Identification Standard [2] that specifies how resources can be identified in the MPEG-21 framework.
  - The http header information that would be provided if the resource was obtained through an http GET request is provided in a `Descriptor/Statement` construct, the content of which is compliant with an XML Schema [http://purl.lanl.gov/STB-RL/schemas/2004-08/HTTP-HEADER.xsd] specifically defined for the mod_oai project.
- The Web resource itself – that is the datastream - is provided in a construct that contains one or two `Resources` in a `Component` that is a child element of the aforementioned `Item`:
  - In all cases, the datastream is provided by-reference by including the URI of the Web resource as the content of the ref attribute of a `Resource` element.
  - In cases where the filesize of the datastream does not exceed a preset and configurable value, the datastream is also provided by-value as the content of a `Resource` element. In this case the datastream is base64 encoded.
  - In both cases, the MIME type of the Web resource is expressed by the mimeType attribute of the `Resource` element.
- The top-level `Item` is embedded in the `DIDL` root element to obtain an XML document that is compliant with MPEG-21 DIDL.

# 5 CRAWLING VS. HARVESTING

During the development of mod_oai, we have encountered a number of subtle but important issues that illustrate the fundamental differences between web crawling and OAI-PMH harvesting. The first of which is that traditional OAI-PMH applications are deterministic with respect to the number of records that the repository holds. Most OAI-PMH repository implementations are accessing a database in which all possible records are knowable. However, web harvesting is different. We define $U$ as the set of all possible URLs for a particular web server, and $F$ as the set of files that the web server can see. Apache maps $U \rightarrow F$, and mod_oai maps $F \rightarrow U$. Neither function is 1-1 nor onto. We can easily check if a single URL maps to $F$, but given $F$ we cannot (easily) generate $U$.

One problem is that Apache can "cover up" legitimate files. Consider two files, A and B, on a web server. Now consider an httpd.conf file with these directives:

```
Alias /A /usr/local/web/htdocs/B
Alias /B /usr/local/web/htdocs/A
```

The URLs obtained by web crawling and the URLs obtained by OAI-PMH harvesting will be in contradiction. Files can also be covered up by Apache's `Location` directive that is used with (among other things) Apache modules. For example, a file named "server-status" would be exported by mod_oai, but would not be accessible directly from the web server if the "mod_status" module is installed. Although it would be expensive, mod_oai could work to resolve all the `Alias` and `Location` httpd.conf directives and could resolve all the symbolic links and all other operations to build a complete view of the files available from `DocumentRoot`. However, there are still side effects that come from the filesystem itself. Figure 2 shows a series of Unix shell commands that shows how user directories are mounted as necessary using the Network File System (NFS). NFS mounted directories shared between many departmental machines is a common deployment scenario and makes it extremely difficult to include `UserDir` files (e.g., http://www.cs.odu.edu/~mln/) in mod_oai responses since there is no (easy) way to know in advance all possible users. However, these files constitute the majority of files accessible from a web server in a shared-user environment such as a university department. mod_oai currently ignores files impacted by `Alias`, `Location` and `UserDir`, although future versions will resolve these conflicts.

```
whiskey.cs.odu.edu:/ftp/WWW/conf% ls /home
liu_x/  mln/
whiskey.cs.odu.edu:/ftp/WWW/conf% ls -d /home/tharriso
/home/tharriso/
whiskey.cs.odu.edu:/ftp/WWW/conf % ls /home
liu_x/  mln/  tharriso/
whiskey.cs.odu.edu:/ftp/WWW/conf %
```

**Fig. 2.** Just-in-time mounting of directories with NFS.

Another problem in web crawling vs. OAI-PMH harvesting is that web servers will frequently transform files before serving them to the client. Some files are dynamic in their nature (e.g, .cgi, .php, .shtml, .jsp, etc.) and it is a potential security-hole to export the unprocessed file. For example, a PHP file might have database passwords in the source file. When accessed directly through the web server, Apache will resolve all the database calls and replaced the PHP code with the appropriate output.

This level of processing is not available to mod_oai. Currently, mod_oai will ignore any file that has requires server-side processing (by checking to see if there is a handler registered for the particular file type). We are currently working on techniques to correctly export dynamic files.

mod_oai currently handles files protected by Apache by checking each file before it is included in a response to see if the necessary credentials in the current http connection are sufficient to meet the requirements specified in the .htaccess file. Since harvesting is not interactive, mod_oai will not prompt for a password if one is required. Rather, any necessary passwords will have to be included in the http environment when the harvest is begun. mod_oai will not advertise any file that the request does not have the credentials to retrieve. This prevents OAI-PMH errors from being generated when a harvester tries to access a protected file, but it does mean that harvesters with different credentials will see a different "view" of the same mod_oai baseURL.

Somewhat related is the problem that Apache will advertise files that it cannot read. For example, a file can be seen in a directory listing, but if the permissions on the file are "000", then no one can actually read the file. The file is listed by Apache since its existence is actually a property of the parent directory, not the file itself. To preserve OAI-PMH semantics, mod_oai will not include such files in responses.

Finally, Apache uses the `IndexIgnore` directive to specify patterns for filenames that should not be included in a directory listing (e.g., "foo~", "foo#" and other file version conventions). However, if requested directly (e.g., http://www.cs.odu.edu/index2.html~), Apache will serve it. The Apache semantics in this scenario are similar to "hidden" files in Unix. However, this has serious implications for OAI-PMH – it would be equivalent to the undesirable scenario where more files are available via GetRecord than ListRecords. To preserve OAI-PMH semantics, mod_oai currently ignores any files of the type specified in `IndexIgnore`.

## 6 EXPERIMENTAL RESULTS

To examine the performance of mod_oai, we compared OAI-PMH harvesting using the OCLC Harvester [25] with the wget web crawling utility [9]. We used a copy of the Old Dominion University Computer Science Department web pages (http://www.cs.odu.edu/) as a testbed. We excluded a number of files, including user file (~user), web mail files and data files from a survey utility. Overall, the testbed includes 5268 files that use 292Mb disk space. The server was at ODU (whiskey.cs.odu.edu, Intel Xeon CPU 2.40GHz, 1.5G RAM) and the client was at LANL (adelie.lanl.gov, Intel Xeon CPU 2.80GHz, dual CPU, 6G RAM).

We performed two experiments. The first uses the departmental homepage as a seed, thus crawled files will be only the ones reachable from the homepage. In the second experiment we created an html list of all files (with "find . -type f") to use as a seed. The significant difference between the numbers of files detected in both experiments (first row of Table 1) is due the reasons discussed in the previous section. The time-stamping in wget is turned on using `--timestamping' (`-N') option. This will cause wget to check whether a local file of the same name exists and only download the remote file if it is "newer" than the local file. Table 1 shows the number of files accessed by wget in both scenarios. Using the "find" seed, downloads more URLs (5739) than there are files (5268). This is because it finds additional URLs that the seed points to, including directories and broken links. The full wget command is:

wget -r --no-parent --exclude-directories=/modoai,~ -N $INDEX -o $INDEXLOG -P $INDEXMIRROR --dns-timeout=1 --connect-timeout=1 --tries=1

**Table 1**. Files accessed by wget.

| | index.html as seed | "find . -type f" as seed |
|---|---|---|
| # of files in baseline | 709 | 5739 |
| # of files in update (25%) | 114 | 1318 |

We performed a baseline with both wget and mod_oai with all file modification dates set to "2000-01-01". For our second test, we touched 25% of the files (1335 files) to make their modification dates "2002-01-01". This simulated the monthly update rate expected for ".edu" sites [7]. For mod_oai, we issued two different requests types (`ListRecords`, `ListIdentifers`), and two different "from" values (1900-01-01, 2001-01-01). We restarted Apache after each round of harvesting was done.

Figure 3 shows the time required for the baseline for all files and Figure 4 shows the time required for just the updated files. It is surprising to see wget takes more time in accessing only the updated files. Apache log file shows that wget uses both the http HEAD and GET methods to check the time. Thus in checking for updates, wget will use more http requests (5739 HEAD + 1318 GET).

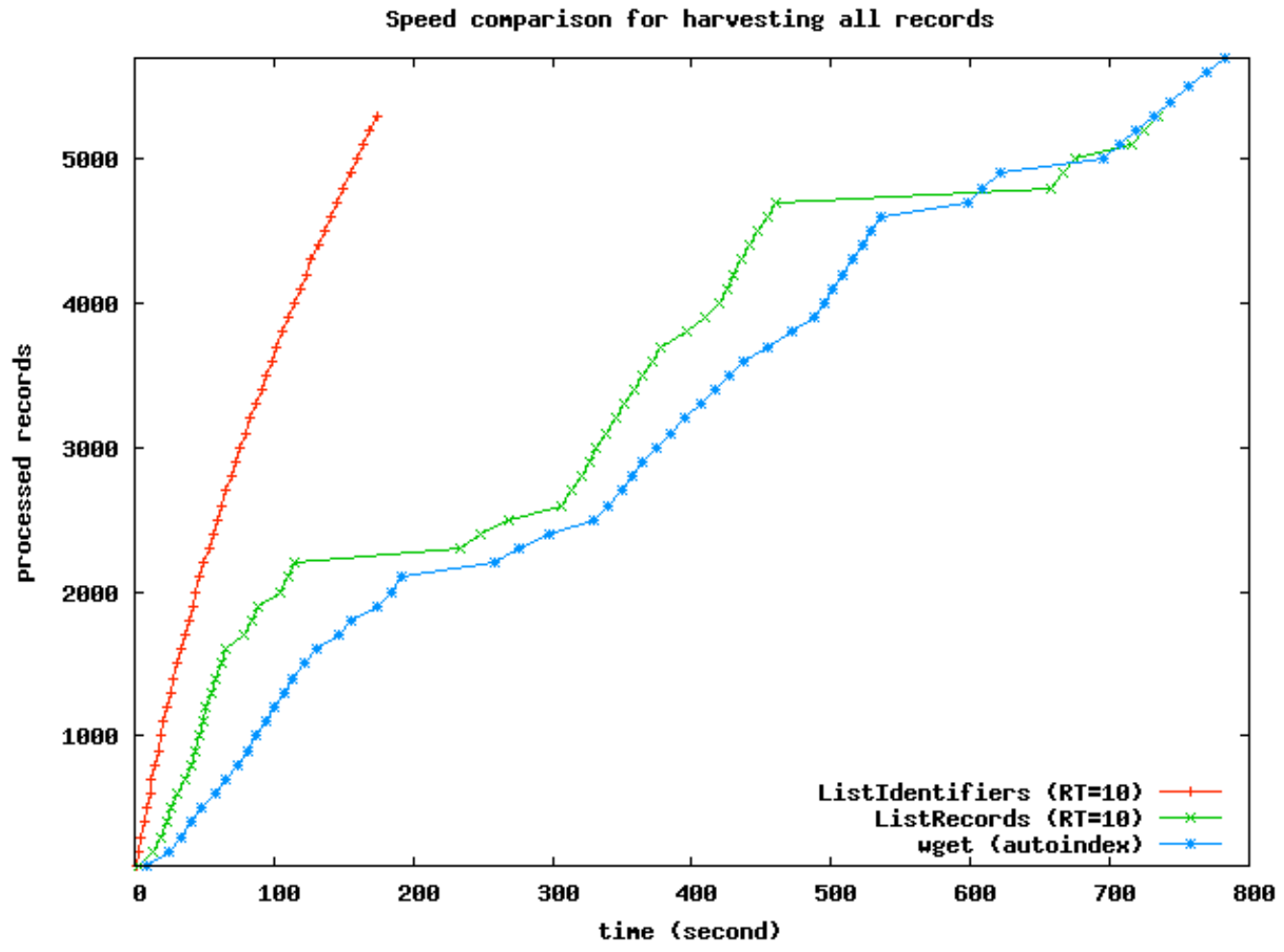

**Fig. 3.** Baseline wget and mod_oai.

We also tested the performance of mod_oai with various resumptionToken sizes (Figure 5). With ListRecords, the performance increased leveled off at a resumptionToken size of 50. With ListIdentifiers, performance continued to increase with increasing the resumptionToken size. This is due to the fact that ListRecords returns the base64-encoded file, and ListIdentifiers returns just the resource identifiers. This suggests that we should have different resumptionToken sizes for ListRecords and ListIdentifiers.

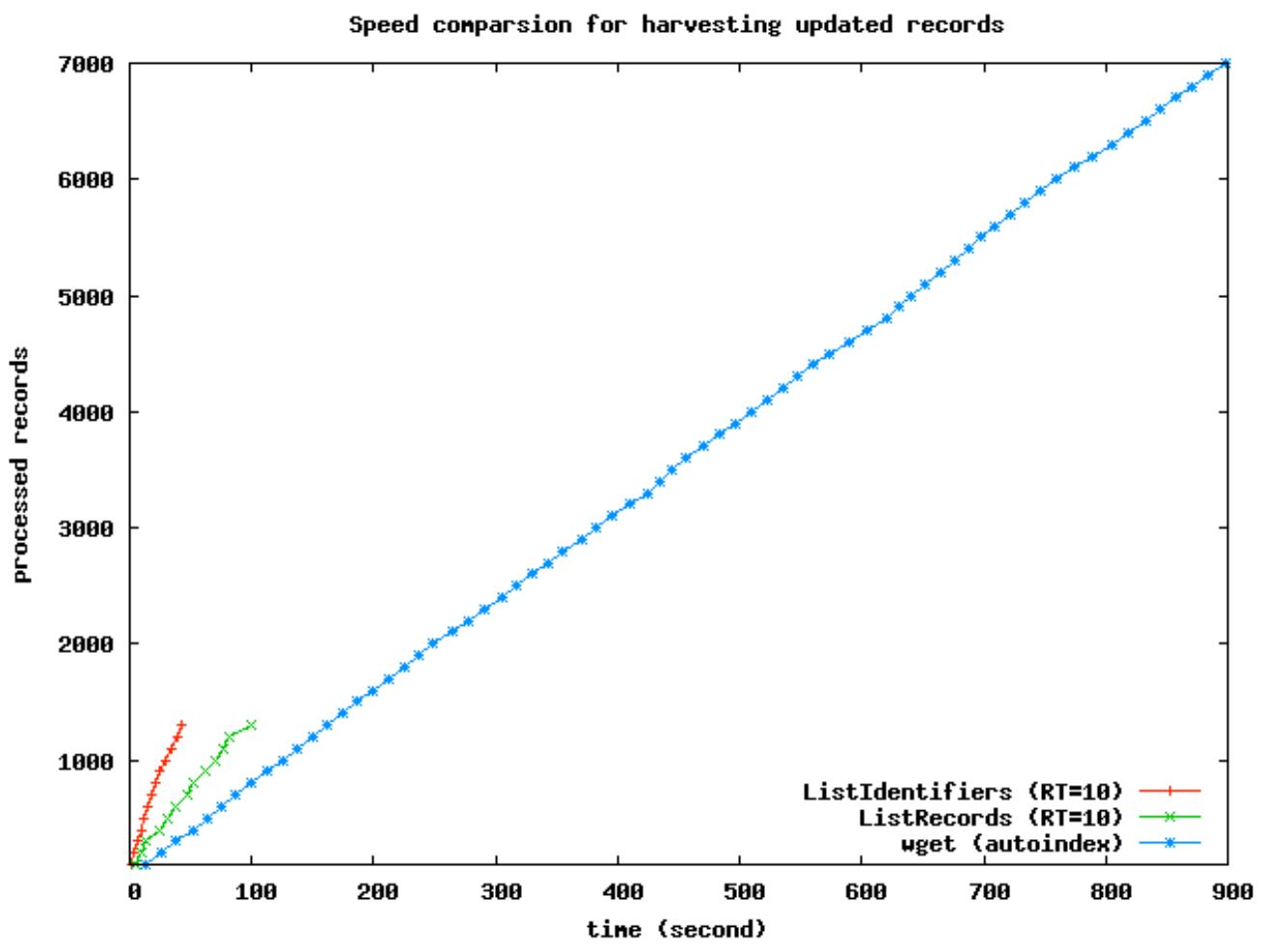

**Fig. 4.** wget and mod_oai after 25% file updates.

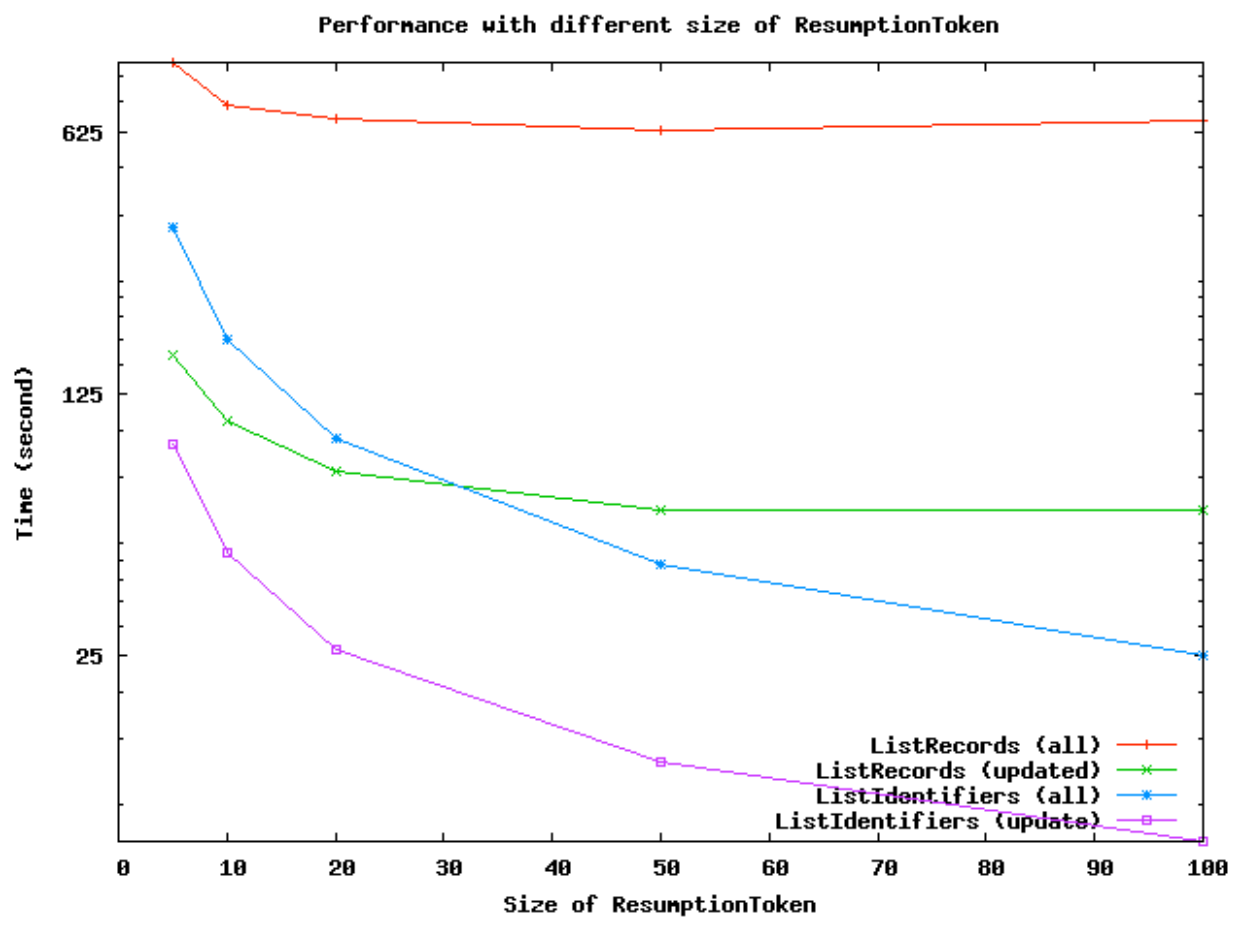

**Fig. 5.** Performance with different resumptionToken sizes.

# 7 CONCLUSIONS

mod_oai is an Apache 2.0 module that exposes a web server as an OAI-PMH repository. We have described how the OAI-PMH semantics are interpreted in the context of an Apache server. We have also listed some of the mismatches that occur between the filesystem, Apache web server and the mod_oai module that complicate the semantic mapping at each level. The most significant of which is that mod_oai currently does not support server-side processed files; we will address this in a future release. We have also presented initial performance results of mod_oai and wget on a typical university department web site. We have shown that mod_oai offers comparable performance to wget for baseline harvests and outperforms wget for when file updates are considered.

mod_oai is not intended to replace existing OAI-PMH repositories, but rather to bring OAI-PMH semantics of incremental harvesting based on datestamps and sets to general web servers. mod_oai can be used to generate a list of URLs for regular web crawlers (using ListIdentifiers), or it can be used to harvest MPEG-21 DIDL encoded versions of the content (using ListRecords).

**Acknowledgements**

mod_oai is supported by the Andrew Mellon Foundation. Aravind Elango contributed to the mod_oai source code. Jeroen Bekaert contributed to the MPEG-21 DIDL profile of mod_oai. The mod_oai web site is www.modoai.org.